\documentclass{article}

\usepackage{arxiv}

\usepackage[utf8]{inputenc} 
\usepackage[T1]{fontenc}    
\usepackage{hyperref}       
\usepackage{url}            
\usepackage{booktabs}       
\usepackage{amsfonts}       
\usepackage{nicefrac}       
\usepackage{microtype}      
\usepackage{lipsum}
\usepackage{algorithmic}
\usepackage{algorithm}
\usepackage{hyperref}
\hypersetup{hidelinks}
\usepackage{multirow}
\usepackage{tikz}
\usepackage{subfigure}
\usepackage{amssymb}
\usepackage{amsmath}
\usepackage{amsthm}
\title{SSD-SGD: Communication Sparsification for Distributed Deep Learning Training}

\author{
  Yemao Xu\\
  Department of Computer Science\\
  National University of Defense Technology\\
  \texttt{xuyemaovip@nudt.edu.cn} \\
   \And
  Dezun Dong\thanks{Corresponding Author} \\
  Department of Computer Science\\
  National University of Defense Technology\\
  \texttt{dong@nudt.edu.cn} \\
  \And 
  Yawei Zhao\\
  Department of Computer Science\\
  National University of Defense Technology\\
  \texttt{yaweizhao@nudt.edu.cn}
  \And
  Weixia Xu \\
  Department of Computer Science\\
  National University of Defense Technology\\
  \texttt{xuweixia@nudt.edu.cn} \\
  \And
  Xiangke Liao \\
  Department of Computer Science\\
  National University of Defense Technology\\
  \texttt{xkliao@nudt.edu.cn} \\
}

\begin{document}
\maketitle

\begin{abstract}
Intensive communication and synchronization cost for gradients and parameters is the well-known bottleneck of
distributed deep learning training. Based on the observations that Synchronous SGD (SSGD) obtains good convergence
accuracy while asynchronous SGD (ASGD) delivers a faster raw training speed, we propose Several Steps Delay SGD (SSD-SGD)  to combine their merits, aiming at tackling the communication bottleneck via communication sparsification.
SSD-SGD explores both global synchronous updates in the parameter servers and asynchronous local updates in the workers in each periodic iteration. The periodic and flexible synchronization makes SSD-SGD achieve good
convergence accuracy and fast training speed. To the best of our knowledge, we strike the new balance between
synchronization quality and communication sparsification, and improve the trade-off between accuracy and training
speed. Specifically, the core components of SSD-SGD include proper warm-up stage, steps delay stage, and our novel
algorithm of global gradient for local update (GLU). GLU is critical for local update operations to effectively
compensate the delayed local weights. Furthermore, we implement SSD-SGD on MXNet framework and
comprehensively evaluate its performance with CIFAR-10 and ImageNet datasets. Experimental results show that
SSD-SGD can accelerate distributed training speed under different experimental configurations, by up to 110\%, while
achieving good convergence accuracy.
	
\end{abstract}


\section{Introduction}
\label{sec:introduction}

Deep Neural Networks (DNNs) have achieved success in many AI tasks, including image recognition~\cite{real2019regularized}, speech recognition~\cite{ kim2017residual}, natural language processing~\cite{karpathy2015visualizing} and so on. However, training a big DNN model is usually time-consuming~\cite{facebookhpca2018}, and distributed training systems are deployed to improve the productivity of training~\cite{chilimbi2014project, xing2015petuum}. The first key problem of distributed training is the \textit{intensive communication} among workers~\cite{asplos2020qian}, and it is crucial to eliminate the potential communication bottleneck~\cite{ICML2019linear}. 

To address the communication challenge, both system and algorithm techniques have been proposed. Regarding the system-level optimization, methods like increasing the batch size~\cite{goyal2017accurate, lars, jia2018highly, sun2019optimizing}, communication scheduling techniques~\cite{p3, tictac, bytescheduler}, deploying faster network fabrics~\cite{tpupod, NVIDIA} and efficient network topologies~\cite{bml, dong2020eflops} are designed. In terms of algorithm-level optimization, works like optimizing the communication algorithms~\cite{baidu, cho2017powerai, sergeev2018horovod}, gradient compression~\cite{wen2017terngrad, alistarh2017qsgd, yu2018gradiveq, lin2017deep, wu2018error} and asynchronous update mechanisms~\cite{hogwild, adadelay, sequence, dcasgd-2017} are also conducted.

The synchronous stochastic gradient descent (SSGD) and asynchronous SGD (ASGD) are two widely used algorithms when performing distributed training. SSGD obtains good convergence accuracy while its training speed deteriorates severely due to the global synchronous communication barrier.
The key to improving the training performance of SSGD is reducing the communication traffic. Therefore, gradient compression techniques received intensive research interests among the optimization fields. On the one hand, these techniques are orthogonal to most of other optimization works. On the other hand, they can cut down the communication overhead dramatically. However, programmers need to adopt some extra operations, such as \textit{Momentum Correction}, \textit{Gradient Clipping}, \textit{Momentum Factor Masking}, to mitigate the accuracy loss problem when using these approaches. 
Besides, programmers need to adjust the value of gradient sparsity in the beginning when using DGC~\cite{lin2017deep}. Layers which are sensitive to quantization operations also call for special treatment throughout the training process. 
The complexity of these great gradient compression techniques limit their wide applications, and we intend to propose a more general strategy to reduce and overlap the communication cost for efficient use of system resources.

ASGD is proposed to break the synchronous barrier to achieve faster training speed while it often suffers from accuracy drop problem because of the weight delay problem~\cite{chen2016revisiting}. Regarding the optimization for ASGD, the key is preserving the convergence accuracy. S. Sra et al.~\cite{adadelay} develop the AdaDelay (Adaptive Delay) algorithm to preserve the global convergence rate for convex optimization. Z. Zhou et al.~\cite{sequence} prove it is possible to obtain global convergence guarantees for ASGD with unbounded delays by using a judiciously chosen quasilinear step-size sequence. Delay compensated ASGD (DC-ASGD)~\cite{dcasgd-2017} makes the optimization behavior of ASGD closer to that of sequential SGD via introducing compensation operations in the master node. Although these ASGD based optimizations achieve better convergence accuracy, it is impossible for them to obtain faster training speed than vanilla ASGD. We intend to design an method to achieve faster training speed than ASGD while ensuring convergence accuracy via communication optimization. 

Based on the features of SSGD and ASGD, the combination of their merits would probably lead to dramatical drop in communication cost.
In this paper, we aim to explore whether it is possible to derive an general scheme for cutting down the communication overhead, thus reducing the iteration time by embracing the advantages of SSGD and ASGD, while maintaining the convergence accuracy without introducing complex operations.

To this end, we first formulate cutting down the average iteration time in distributed DNN training as an optimization problem.
Then we propose SSD-SGD, which combines the merits of SSGD and ASGD, to minimize the average iteration time by reducing part of the communication operations, or communication sparsification. The intrinsical motivation of SSD-SGD is reducing the communication cost and increasing the overlap between computation and communication, achieving good convergence accuracy and fast training speed at the same time. 
We implement SSD-SGD in MXNet~\cite{chen2015mxnet} and conduct experiments on various DNN models to evaluate its convergence performance and training speed. Experimental results show that SSD-SGD accelerates DNN training speed by up to 110\% while achieving good convergence accuracy.
In summary, the major contributions of our work are as follows:

\begin{itemize}
	
	\item We propose to reduce part of the communication operations in several consecutive iterations to reduce the average iteration time and we formulate an optimization problem for this.
	
	\item We come up with SSD-SGD to minimize the average time via communication sparsification without affecting the convergence accuracy, and we implement a prototype of SSD-SGD on MXNet.
	
	\item We comprehensively evaluate the convergence accuracy and training speed improvement of SSD-SGD, and experimental results demonstrate its promising benefits in distributed DNN training.
\end{itemize}

The rest of this paper is organized as follow.
We present some preliminaries and formulation of problem we need to tackle in Section~\ref{sec:preliminaries}, and we propose a solution named SSD-SGD to the problem as well as its implementation in Section~\ref{sec:solution}. We then evaluate the performance of SSD-SGD in Section~\ref{sec:eval_set}. Section~\ref{sec:relate} introduces the related work and we conclude in Section~\ref{sec:conclusion}.

\section{Preliminaries and Problem Formulation}
\label{sec:preliminaries}

We introduce some preliminaries and formulate the problem in this section. For the ease of presentation, Table~\ref{tab:variable-des} lists the frequently used notations throughout this paper.

\begin{table}[htbp]\scriptsize
	\renewcommand{\arraystretch}{1.3}
	\caption{The involved variables and their definitions.}
	\vspace{-0.5em}
	\label{tab:variable-des}
	\centering
	\begin{tabular}{|c|c|}		
		\toprule
		Name & Description \\
		\hline
		$T_{iter}$ & Training time in one iteration.\\
		\hline
		$T_{f}$ & Forward computation cost in one iteration. \\
		\hline
		$T_{b}$ & Backward computation cost in one iteration. \\
		\hline
		$T_{c}$ & Communication cost in one iteration. \\
		\hline
		$h_{b}^{j}$ & Time of the backward computation of layer $j$. \\
		\hline
		$h_c^{j}$ &  Time of communication of layer $j$\\
		\hline
		$h_{s}^{j}$ & Time of sending gradient of layer $j$\\
		\hline
		$h_{loc}^{j}$ &  Time of local update of layer $j$\\
		\hline
		$w_{t,i}^{'}$ & Local weight in the $i^{th}$ worker \\
		\hline
		$pre\_weight$ & Previous retrieved weight from the servers \\
		\hline
		$loc\_lr$ &  Local learning rate in workers \\
		\hline
		$grad_{sync}^{t,i}$ & Calculated global gradient for local update\\
		$(grad_{sync})$  & in the $i^{th}$ worker \\
		\hline
		$grad_{t,i}^{'}$ & Local gradient in the $i^{th}$ worker\\
		\hline
		$w_{loc}^{m}$ & Updated local weight in the $m^{th}$ device \\
		\hline
		$grad_{t}$ & Averaged gradient in parameter servers\\
		\hline
		$lr$ & Global learning rate in parameter servers \\ 
		\hline
		$w_{t}$ & Global weight in parameter servers \\ 
		\hline
		$wd$ & Weight decay in parameter servers \\ 
		\hline
		$mom_{t}$ & Momentum in parameter servers \\ 
		\bottomrule
	\end{tabular}
\end{table}

\subsection{Distributed DNN Training}

\begin{figure}[b]
	\centering
	\includegraphics[width= 3.2 in]{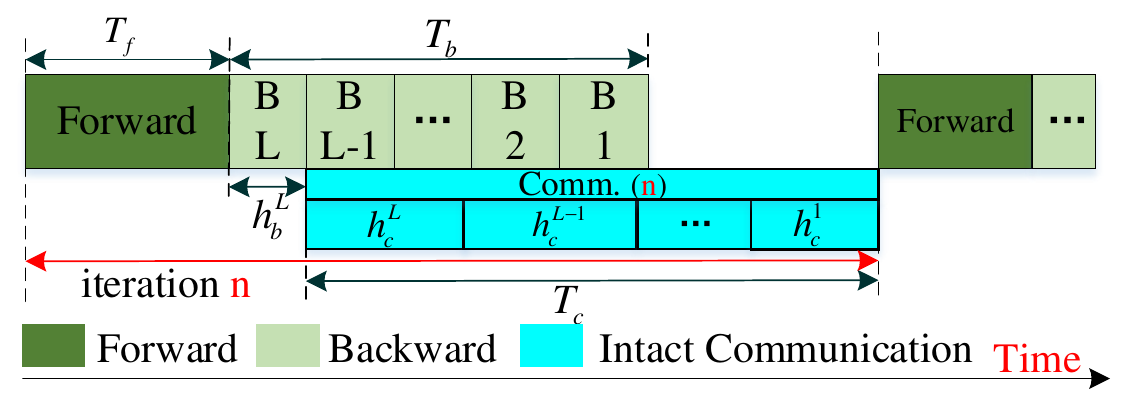}
	\caption{Training process under SSGD ($\sum_{i=1}^{L-1}h_{b}^{i} \leq \sum_{j=2}^{L}h_{c}^{j}$).}
	\label{fig:sgd}
\end{figure}

DNN training are typically conducted with SGD, and for an optimization problem of the form 
$min_{w\in{R^d}}\frac{1}{N}\sum_{j=1}^{N}f_{j}(w)$, where $w\in{R^d}$ is the parameters of the neural network and $f_{j}: R \rightarrow R^{d}$ is the loss function of the $j^{th}$ training example, the distributed SGD update for $K$ workers in the $t^{th}$ iteration is given as

\begin{equation}
\label{eqn:sgd-update}
w_{t+1} = w_{t} + lr[\frac{1}{KB}\sum_{i=1}^{K}\sum_{j \in \xi_{(t)}^{i}}\triangledown f_{j}(w_{t})]
\end{equation} 

where $lr$ is the global learning rate and $\xi_{(t)}^i$ denotes the mini-batch of training data in the $i^{th}$ worker, with the mini-batch size $B$. 

By the end of each iteration, gradients of all workers are aggregated firstly and the parameters are updated with the averaged gradients following Equation~\ref{eqn:sgd-update}. Therefore, in addition to the forward and backward computation overhead, the communication overhead also plays an important role in the training process. 
The overhead for each iteration ($T_{iter}$) can be divided into three parts, forward computation cost ($T_{f}$), backward computation cost ($T_{b}$) and communication cost ($T_{c}$). Based on the sum of backward computation cost ($\sum_{i=1}^{L-1}h_{b}^{i}$) and sum of communication overhead ($\sum_{j=2}^{L}h_{c}^{j}$), the iteration time can be expressed in two cases, as shown in Equation~\ref{eqn:iter}, where $L$ is the layer number of the model, and Fig.~\ref{fig:sgd} illustrates the second case. What should be emphasized is that there may exist bubble between two adjacent layers in the communication process, and we assign the cost introduced by the bubble to the communication cost of the former layer.

\begin{equation}
\label{eqn:iter}
T_{iter}=\left\{
\begin{array}{lr}
T_{f} + T_{b} + h_{c}^{1}  & \sum_{i=1}^{L-1}h_{b}^{i} \textgreater \sum_{j=2}^{L}h_{c}^{j}\\
\\
T_{f} + T_{c} + h_{b}^{L} &  \sum_{i=1}^{L-1}h_{b}^{i} \leq \sum_{j=2}^{L}h_{c}^{j}
\end{array} 
\right.
\end{equation}

\begin{figure*}
	\centering
	\begin{minipage}{0.7\linewidth}
		\centering
		\includegraphics[width= 5.25 in, height= 1.2in]{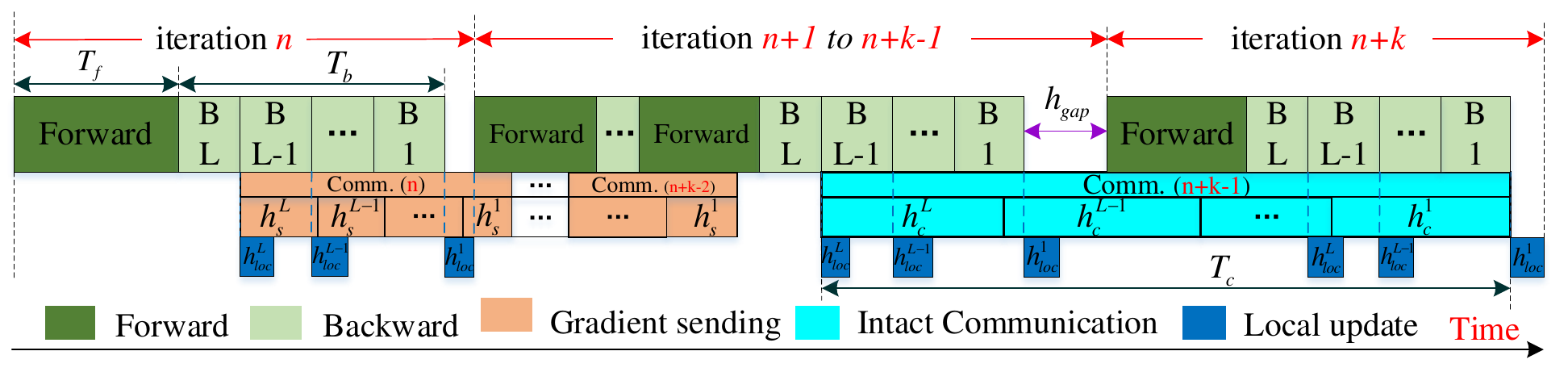}
		\centerline{(a) Case one: $\sum_{j=1}^{L}h_{s}^{j} $ \textless $T_{b}+T_{f}+h_{loc}^{1}$}
		\parbox{6.5cm}{\small \hspace{1.5cm}}
	\end{minipage}
	\vfill
	\begin{minipage}{0.7\linewidth}
		\centering
		\includegraphics[width= 5.25 in, height= 1.2in]{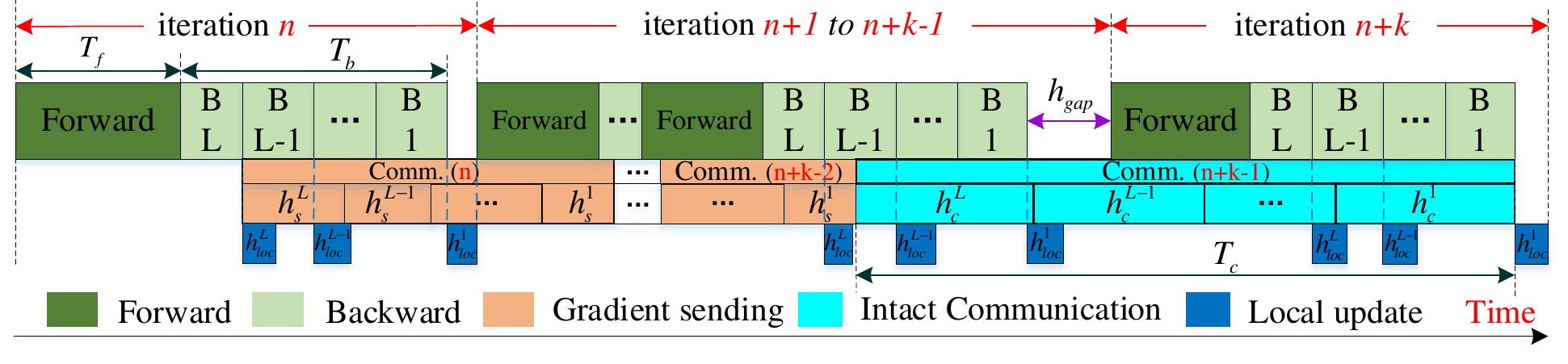}
		\centerline{(b) Case two: $\sum_{j=1}^{L}h_{s}^{j}$ $\geq$ $T_{b}+T_{f}+h_{loc}^{1}$}
		\parbox{6.5cm}{\small \hspace{1.5cm}}
	\end{minipage}	
	\centering
	\caption{Two cases of training process for $k$ successive iterations with communication sparsification.}
	\label{fig:k-iter}
\end{figure*}

The communication cost for the $j^{th}$ layer can be further partitioned into four different overheads, namely the gradient sending (i.e. \textit{Push}) cost ($h_{s}^{j}$), weight retrieving (i.e.\textit{Pull}) cost ($h_{r}^{j}$), synchronous cost ($h_{sync}^{j}$) and update cost ($h_{up}^{j}$). Then we have

\begin{equation}
\label{eqn:tc}
h_{c}^{j} = h_{s}^{j} + h_{r}^{j} + h_{sync}^{j} + h_{up}^{j}
\end{equation}

\subsection{Problem Formulation}

Without improvements in the computing capability of hardwares, $T_{f}$ and $T_{b}$ keep almost unchanged throughout the training process.
When $\sum_{i=1}^{L-1}h_{b}^{i} \textgreater \sum_{j=2}^{L}h_{c}^{j}$, the iteration time $T_{iter}$ mainly depends on the computation overhead $T_{f}$ and $T_{b}$, it does not make great sense to optimize the communication cost $h_{c}^{1}$. When $\sum_{i=1}^{L-1}h_{b}^{i} \leq \sum_{j=2}^{L}h_{c}^{j}$, the key to reducing the training cost $T_{iter}$ is cutting down $T_{c}$, and this paper focus on the second case to optimize the communication cost.

Considering the communication cost $T_{c}$ vibrates slightly in one single iteration without the gradient compression techniques, one way for optimizing the training cost is reducing the involved communication operations like $Push$ and $Pull$. However, eliminating the same communication operation for every iteration will lead to divergence of the model. Therefore, we intend to minimize the average iteration time of $k (k \textgreater1)$ iterations by reducing part of communication operations involved in the $k$ iterations.

As the synchronous barrier deteriorates the training speed severely, we choose to eliminate the $Pull$ operation to break this barrier, and the $Push$ operation is kept to maintain the convergence accuracy.
The $Pull$ operation is executed only once every $k$ iterations, and Fig.~\ref{fig:k-iter} shows the training process, the blue blocks denote the $local$ $update$ operations, which are performed by local workers to update their own weights to start the following iteration shortly. In the first $k$-$1$ iterations, the $Pull$ operations are removed, then workers do not need to wait for the updated parameters, and the synchronous cost $h_{sync}^{j}$ is reduced. The parameter update cost $h_{up}^{j}$ is also eliminated as the update operations are conducted by the parameter servers.
Under this circumstance, the communication cost consists of only the $h_{s}^{j}$, and it can be largely overlapped by the computation overhead. In the $k^{th}$ iteration (iteration $n+k-1$), workers need to pull back the updated weights from the servers, four kinds of overheads are involved in the communication process. However, training task of iteration $n+k$ can still be launched directly before retrieving the update weights. 
The retrieved parameters will be used for the $local$ $update$ operations in iteration $n+k$. It is noteworthy that $local$ $update$ operations can be parallelized with the $Push$ operations as both kinds of operations only have \textit{read-dependency} on the calculated gradients.





\begin{equation}
\label{eqn:k-relation}
T_{avg,k}=\left\{
\begin{array}{lr}
T_f+T_b+h_{loc}^{1}+\frac{1}{k}(T_c+h_{b}^{L}-T_f-2T_b-h_{loc}^{1})\\
(Case 1: \sum_{j=1}^{L}h_{s}^{j} \textless T_{b}+T_{f}+h_{loc}^{1}) \\
\\
\sum_{j=1}^{L}h_{s}^{j}+\frac{1}{k}(T_c+h_{b}^{L}-T_b-\sum_{j=1}^{L}h_{s}^{j}) \\
(Case 2: \sum_{j=1}^{L}h_{s}^{j} \geq T_{b}+T_{f}+h_{loc}^{1})
\end{array}
\right.
\end{equation}

Equation~\ref{eqn:k-relation} presents the average iteration time for $k$ iterations, Case 1 and Case 2 correspond to the two cases in Fig.~\ref{fig:k-iter}, where the computation process of iteration $n+k$ completes before the communication process of iteration $n+k-1$ finishes.
For situations where the computation process of iteration $n+k$ ends later (Case 3), the average iteration time equals $T_f + T_b + h_{loc}^{1}$ and it is not related to iteration number $k$. Case 3 is not the optimization target in this paper as computation cost dominates the training process.

Case 1: As the computation process of iteration $n+k$ finishes earlier than the communication process of iteration $n+k-1$, value of $h_{gap}$ is larger than $h_{loc}^{1}$, and value of $T_c+h_{b}^{L}$ equals $T_f+2T_b+h_{gap}$. Therefore, we can draw the conclusion that $(T_c+h_{b}^{L}-T_f-2T_b-h_{loc}^{1}) \textgreater 0$. Actually, the computation process of iteration $n+k$ is started after the $local$ $update$ operation for the first layer in iteration $n+k-1$, we move it back to be align with the communication process for the clarity of calculating the average iteration time. Based on the above conclusion, a larger $k$ will contribute to smaller average iteration time.

Case 2: $\sum_{j=1}^{L}h_{s}^{j}$ covers only the sum of $gradient$ $sending$ overhead in one iteration, thus $T_c$ is more than 2 times the value of $\sum_{j=1}^{L}h_{s}^{j}$, and $(T_c+h_{b}^{L}-T_b-\sum_{j=1}^{L}h_{s}^{j}) \textgreater (\sum_{j=1}^{L}h_{s}^{j}+h_{b}^{L}-T_b)$. 
Case 2 corresponds to condition $\sum_{j=1}^{L}h_{s}^{j}$ $\geq$ $T_{b}+T_{f}+h_{loc}^{1}$, we can draw the conclusion that $(T_c+h_{b}^{L}-T_b-\sum_{j=1}^{L}h_{s}^{j}) \textgreater 0$ and a larger $k$ will also lead to smaller average iteration time.

Smaller iteration time means faster training speed and a larger $k$ results in better training speed. However, the convergence accuracy should also be considered in the training process as a larger $k$ will probably lead to more serious weight delay problem, imposing negative influence on the convergence accuracy. Our target is to maximize the value $k$ while maintaining the convergence accuracy, and we come up with a solution from the experimental and theoretical perspective.

\subsection{Our Design with Communication Sparsification}
Catering to the mechanism above, we come up with SSD-SGD to maximize the value $k$ and maintain the convergence accuracy. In SSD-SGD, each local worker $i \in {[K]}$ evolves a local model by performing local updates using Equation~\ref{eqn:worker} at each iteration, and all workers perform the local update operations asynchronously with the specified algorithm, SGD is used in Equation~\ref{eqn:worker}. The gradients calculated using local weights at each iteration are sent to the servers and servers conduct synchronous global updates following Equation~\ref{eqn:server}. Different from SSGD, the global updated weights are retrieved only once every $k$ iterations in SSD-SGD. Under this circumstance, the local weight $w_{t-1, i}^{'}$ is overwritten by the retrieved global weight $w_{t-1}$ every $k$ iterations, and the communication cost can be reduced significantly via omitting $k$-$1$ times of $Pull$ operations.

\begin{equation}
\label{eqn:worker}
w_{t, i}^{'} = w_{t-1, i}^{'} + lr_{loc}[\frac{1}{B}\sum_{j \in \xi_{(t-1)}^{i}}\triangledown f_{j}(w_{t-1,i}^{'})]
\end{equation}

\begin{equation}
\label{eqn:server}
w_{t} = w_{t-1} + lr[\frac{1}{KB}\sum_{i=1}^{K}\sum_{j \in \xi_{(t-1)}^{i}}\triangledown f_{j}(w_{t-1, i}^{'})]
\end{equation}

For $k$ sequent iterations, the total iteration time of SSGD is $k(T_f+T_c+h_b^L)$, the average iteration time of SSD-SGD is shown in Equation~\ref{eqn:k-relation}. Then we have

\begin{equation}
\label{eqn:ssgd-ssd-sgd}
\triangle T_{k}=\left\{
\begin{array}{lr}
Case1: (k-1)(T_c-T_b+h_b^{L}-h_{loc}^{1})+(T_f+T_b) \\
\\
Case2: kT_f+\frac{k-1}{2}T_c+(k-1)h_b^L+T_b
\end{array}
\right.
\end{equation}

Where $\triangle T_k = T_{iter,k}-T_{iter,k}^{'}$, $T_{iter,k}$ and $T_{iter,k}^{'}$ correspond to the iteration cost for $k$ sequent iterations of SSGD and SSD-SGD, respectively. Value of $\sum_{j=1}^{L}h_{s}^{j}$ is set to $\frac{1}{2}T_c$ for simplicity when calculating $\triangle T_k$. From Equation~\ref{eqn:ssgd-ssd-sgd} we can notice SSD-SGD achieves more performance improvement with a larger $k$ when compared to SSGD.

\section{Implementation and Proof of SSD-SGD}
\label{sec:solution}

In this section, we discuss some details about SSD-SGD. We firstly introduce its core components (1) warm-up stage, (2) local update operation, and (3) steps delay stage. Then we provide the algorithm description and analyze the convergence rate. Finally, we briefly present the implementation process.

\subsection{Warm-up Stage}
The warm-up stage is introduced to make the weights and gradients more stable before the application of steps delay mechanism, value $k$ corresponds to the number of delay steps. Warm-up stage here means training with SSGD in the beginning, which differs from the warm-up stage discussed in~\cite{goyal2017accurate}. To tell their distinction, the former is marked as \textit{warm-up} stage while the latter is marked as $WP$ stage.
The length of warm-up stage is much more like an empirical trick 
and we provide length sensitivity analysis from the experiment perspective in Section~\ref{sec:warm-up sensitivity}.

\subsection{Local Update Operation}
\label{sec:local updater}

Local update operation is introduced to maintain the convergence accuracy. 
Under the $k$ delay steps stage, local workers will retrieve the global weights every $k$ iterations, and workers need to conduct local updates with the GLU algorithm at each iteration.

\subsubsection{GLU Algorithm}
\label{sec:glu-algorithm}

\mbox{GLU} algorithm is applied in local update operation and Equation~\ref{eqn:glu-update} shows its update rule, where 
$w_{t,i}^{'}$ is the local weight in the $i^{th}$ worker,
$lr_{loc}$ is the local learning rate in workers,
$\alpha$ and $\beta$ are two coefficients used to determine the ratio of local and global gradients,
value of $w_{t-1,i}^{'}$ depends on whether local workers perform \textbf{\textit{Pull}} operation, if executed, $w_{t-1,i}^{'}$ equals to global weight $w_{t-1}$ in the parameter sever, otherwise it is the updated local weight.
$grad_{sync}^{t-1,i}$ is the global gradient in the $i^{th}$ worker, calculated using the global weights retrieved from the parameter servers. For simplicity, we refer to $grad_{sync}^{t-1,i}$ as $grad_{sync}$ in the following paper.

\begin{equation}
\label{eqn:glu-update}
w_{t,i}^{'} = w_{t-1,i}^{'} + lr_{loc} * (\alpha*grad_{t-1,i}^{'} + wd*w_{t-1,i}^{'} + \beta*grad_{sync}^{t-1,i})
\end{equation}

The keypoint of GLU algorithm is figuring out the global gradient $grad_{sync}$. Take one step delay as an example, the parameter server update weights with momentum SGD, $w_{t-1}$ and $w_{t-2}$ are global weights maintained in the parameter server, so we have

$$
\begin{aligned}
w_{t-1}  &= w_{t-2} + mom_{t-2}
\end{aligned}
$$
Where $mom_{t-2}$ is the \textit{momentum} in the parameter server and according to its update rule, we have

$$
\begin{aligned}
mom_{t-2} &= -lr*(grad_{t-2} + wd*w_{t-2}) + m*mom_{t-3}
\end{aligned}
$$

Where $m$ is the coefficient of $momentum$, then we have

$$
\begin{aligned}
(1-lr*wd)*w_{t-2} - w_{t-1} &= lr*grad_{t-2} - m*mom_{t-3}
\end{aligned}
$$

After marking $((1 -lr*wd)*w_{t-2} - w_{t-1})$ as $w_{minus}$ and supposing  $grad_{t-2} = grad_{t-3} = ... = grad_{0} = grad_{sync}$, which is reasonable after the warm-up stage. We have

$$
\begin{aligned}
w_{minus} &= lr*\frac{grad_{sync}*(1-m^{t-1})}{(1-m)} 
\end{aligned}
$$

As $m < 1.0$, after training for hundreds of iterations (the warm-up stage), we have 

$$
\begin{aligned}
grad_{sync} &= \frac{w_{minus}*(1-m)}{lr} \approx  \frac{(w_{t-2} - w_{t-1})*(1-m)}{lr}
\end{aligned}
$$

Some calculation steps are omitted due to the limited pages, and $grad_{sync}$ is the approximate value of global gradient used for directing the training. To calculate $grad_{sync}$, $w_{t-2}$ and $w_{t-1}$ are stored in the $pre\_weight$ and $w_{t,i}^{'}$ variables respectively in local workers, $pre\_weight$ is used to keep global weight pulled back from the parameter server last time.

\subsection{Steps Delay Stage}
\label{k-steps}
When $k$ steps delay mechanism is used, if $w_{t,i}^{'}$ corresponds to global weight $w_{t-1}$, then $pre\_weight$ corresponds to global weight $w_{t-k-1}$. Thus we have 

$$
\begin{aligned}
grad_{sync} &=  \frac{(w_{t-1-k} - w_{t-1})*(1-m)}{lr*k} \\
&=  \frac{(pre\_weight - w_{t,i}^{'})*(1-m)}{lr*k}
\end{aligned}
$$

After calculating $grad_{sync}$, value of $pre\_weight$ is overwritten by $w_{t,i}^{'}$, or $w_{t-1}$. In the next $k$ iterations and before the retrieve of $w_{t-1+k}$, $w_{t,i}^{'}$ is the local weight, different workers have their unique versions, each worker calculates the $grad_{sync}$ with their own local weights and the same $pre\_weight$.

\subsection{Algorithm Description}
\label{sec:ssd-sgd}

We implement SSD-SGD on MXNet, thus the training process involves update rules of parameter servers and workers. The process of parameter server keeps unchanged when compared to SSGD and we do not present its algorithm with pseudo-code.

\begin{algorithm}[t] 
	\caption{Update rules of local $worker_{i}$} 
	\label{worker-update} 
	\begin{algorithmic}[1] 
		\STATE \textbf{Input}: {warm-up iteration number $wp$, steps delay number $k$}
		\STATE \textbf{Initialize}: {Pull initialized global weight $w_{0}$ from parameter server, iteration number $num$ = 0, time sequence $t$ = 0}
		\WHILE{$num <= wp$}
		\STATE Broadcast $w_{t}$ to local devices ($w_{loc}^{m}$)
		\STATE Computer gradient $grad_{t,i}^{'}$
		\STATE Push $g_{t,i}^{'}$ to the parameter server
		\STATE $t \leftarrow t + 1$
		\STATE Pull $w_{t}$ from the parameter server
		\IF{$num == wp$}
		\STATE $w_{t,i}^{'} = w_{t}$
		\STATE Broadcast $w_{t}$ to local devices ($w_{loc}^{m}$)
		\ENDIF	
		\STATE $num \leftarrow num + 1$
		\ENDWHILE
		
		\REPEAT
		\STATE Compute gradient $grad_{t,i}^{'}$
		\STATE Push $grad_{t,i}^{'}$ to parameter server
		\STATE Update local weight $w_{t,i}^{'}$ with local update optimizer $GLU$
		\STATE $t \leftarrow t+1$
		\IF{$num \% k == k-1$}
		\STATE Broadcast $w_{t,i}^{'}$ to local devices ($w_{loc}^{m}$)
		\STATE Pull $w_{t}$ from the parameter server and copy to $w_{t,i}^{'}$
		\ELSE
		\STATE Broadcast $w_{t,i}^{'}$ to local devices ($w_{loc}^{m}$)
		\ENDIF
		\STATE $num \leftarrow num + 1$
		\UNTIL{forever}
	\end{algorithmic} 
\end{algorithm}

Algorithm~\ref{worker-update} presents the update rules of local workers. Value of $(1+wp)\%k$ should be $0$. In the warm-up stage, each worker broadcasts the retrieved global weight ($w_{t}$) to local devices ($w_{loc}^{m}$) for calculating the gradient, then merge the gradients via a $Reduce$ operation and push the merged gradient to the parameter server. In the last iteration of warm-up stage, the workers make a copy of the pulled weight and keep it in variable $w_{t,i}^{'}$, then broadcast $w_{t}$ to start the next iteration, which is also the first iteration of the $k$ steps delay mechanism. After applying $k$ steps delay strategy, local workers will perform $local$ $update$ and $Push$ operations every iteration (Line $17-18$), these two types of operations can be parallelized as no data dependency exists between them. 
Different from $Push$ operations, $Pull$ requests are sent to parameter servers every $k$ iterations.
Value of the retrieved global weight $w_{t}$ will be assigned to $w_{t,i}^{'}$ (Line $20-25$). In addition, the $Broadcast$ and $Pull$ operations in Line $21$ and $22$ can also be parallelized.
When local workers do not need to send $Pull$ requests, the updated $w_{t,i}^{'}$ are broadcast directly to start the following training task. 

Algorithm~\ref{GLU-opt} shows the process of GLU algorithm. $grad_{sync}$ is calculated with $pre\_weight$ and $w_{t,i}^{'}$, then $grad_{sync}$ and $grad_{t,i}^{'}$ are used for local update. The $pre\_weight$ is overwritten by $w_{t,i}^{'}$ every $k$ iterations. As the retrieved parameters in iteration $i$ will be used for local update in iteration $i+1$, value of $loc\_update\%k$ should be $0$.

\begin{algorithm}[t] 
	\caption{Update rules of optimizer GLU} 
	\label{GLU-opt} 
	\begin{algorithmic}[1] 
		\STATE \textbf{Input}: {steps delay number $k$}
		\STATE \textbf{Initialize}: {Local update operation times $loc\_update$ = 0, $pre\_weight$ is initialized with $w_{t,i}^{'}$ for the first local update operation}
		\STATE Calculate $grad\_sync$ with $pre\_weight$ and $w_{t,i}^{'}$
		\IF{($loc\_update > 0$) and ($loc\_update \% k == 0$)}
		\STATE $pre\_weight = w_{t,i}^{'}$
		\ENDIF
		\STATE Calculate $w_{t+1,i}^{'}$ with $w_{t,i}^{'}$, $grad_{sync}$ and $g_{i}^{t}$
		\STATE $loc\_update \leftarrow loc\_update + 1$
	\end{algorithmic} 
\end{algorithm}

\subsection{Embedding in Framework}

When compared to SSGD in implementation, no modification is needed in the parameter server side. 
For the worker side, SSD-SGD calls for a copy of the retrieved global weights, and we 
store the backup model weights ($pre\_weight$) in the CPU-side memory to avoid occupying the memory of devices (GPU, TPU etc.) used for computing task. SSD-SGD also has extra computational requirement on the local workers to update $w_{t,i}^{'}$, we propose the GLU algorithm and the additional computation only introduces a lightweight overhead. Besides, the $local$ $update$ overhead can be largely overlapped by the $Push$ operation as these two operations can be parallelized.
In addition, we need to define the local updater and adjust the execution sequences of involved operations in steps delay stage.

\textbf{Local Updater.} 
When implementing SSD-SGD, we introduce a new function named $set\_updater\_$ to define the local updater and users can choose the specified algorithm through the  $”-optimizer\_local”$ option when launching the distributed training task.
The GLU algorithm is used in our experiments and we need to modify the $optimizer.py$ file to define it. To achieve better training performance, the computation operations are implemented in $C$++ instead of $Python$, thus we also need to modify the $optimizer\_op-inl.h$, $optimizer\_op.cc$ and $optimizer\_op.cu$ files to define the computation functions.

\begin{figure}[t]
	\centering
	\begin{minipage}{0.4\linewidth}
		\centerline{\includegraphics[width=2 in, height= 1.5 in]{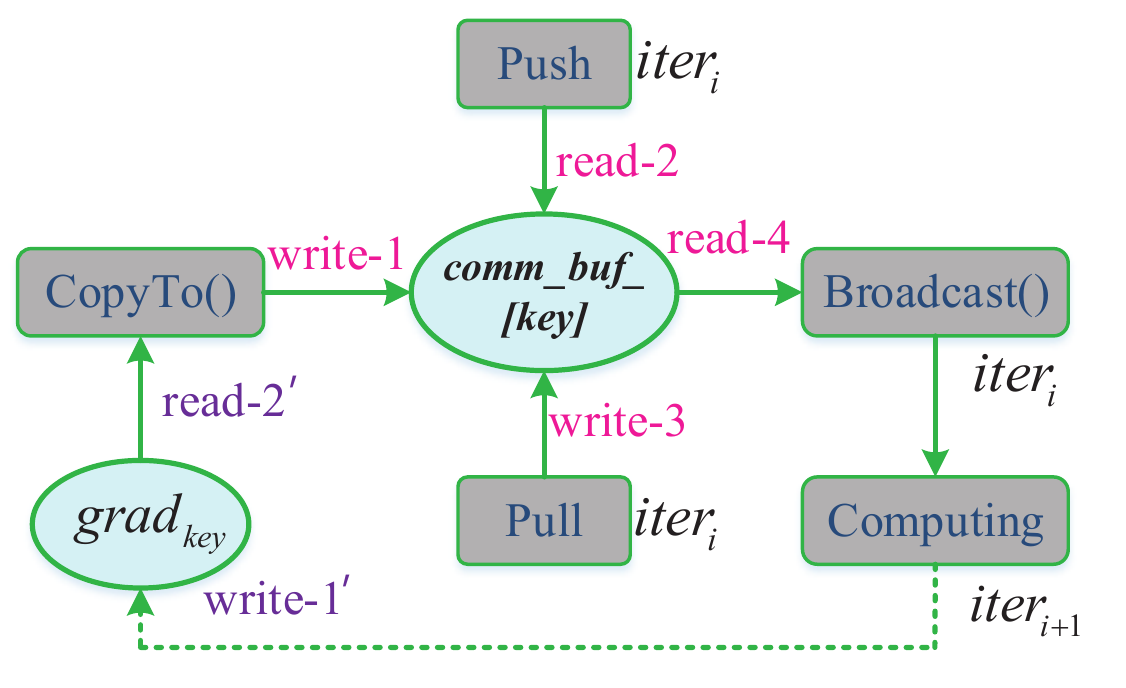}}
		\centerline{(a)}
	\end{minipage}
	\begin{minipage}{0.5\linewidth}
		\centerline{\includegraphics[width=2.4 in, height= 1.5 in]{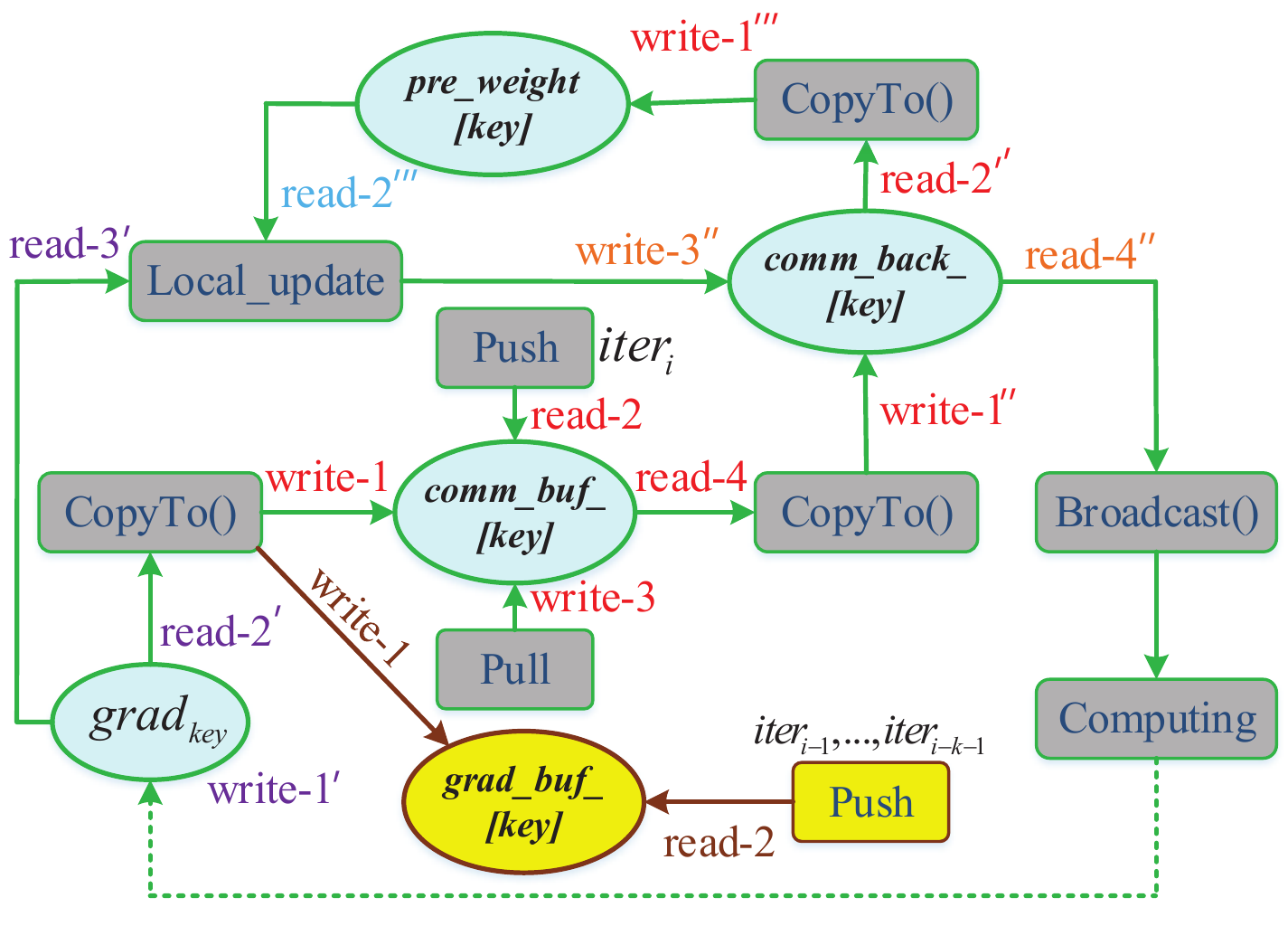}}
		\centerline{(b)}
	\end{minipage}
	\caption{Training procedures of SSGD and SSD-SGD. Numbers at the end of $read$ and $write$ operations describe the order in which they are performed.}
	\label{fig:ssgd-ssd}
\end{figure}

\textbf{Execution Sequences of Operations.} In the warm-up stage, the cluster performs training task under SSGD mechanism and Fig.~\ref{fig:ssgd-ssd}(a) illustrates the original training procedures of SSGD method. $comm\_buf\_[key]$ is the shared variable of $Push$ and $Pull$ operations, leading to data dependency between them, $key$ is index of the corresponding parameter. The reduced gradient ($grad_{key}$) is copied to $comm\_buf\_[key]$ through $\color{magenta}{write-1}$ firstly, then $Push$ operation read the gradient from $comm\_buf\_[key]$ via $\color{magenta}{read-2}$ and send it to the parameter server(s). The $Pull$ operation cannot be executed until the completion of $Push$ operation, it pulls back the updated global weight from the parameter server and store it in $comm\_buf\_[key]$ by $\color{magenta}{write-3}$. Finally the value in $comm\_buf\_[key]$ is broadcast to local devices for computing task of the next iteration. 
Fig.~\ref{fig:ssgd-ssd}(b) presents the training process of SSD-SGD under the steps delay mechanism. The $Pull$ operation pulls back the weight and keeps it in $comm\_buf\_[key]$, then $CopyTo()$ function copies the value to $comm\_back\_[key]$. The $pre\_weight[key]$ is overwritten by $comm\_back\_[key]$ every $k$ iterations and it is used for calculating $grad_{sync}$. The reduced gradient will overwrite the $comm\_buf\_[key]$ or $grad\_buf\_[key]$ for $Push$ operation and also be used for $Local\_update$ operation.
$grad\_buf\_[key]$ is used in iterations with only $Push$ operations and no $Pull$ operations. 
Ultimately, the updated value in $comm\_back\_[key]$ is broadcast for launching the next iteration task. What should be noticed is that operations in red color ($\color{red}{write-1}$, $\color{red}{read-2}$, $\color{red}{write-3}$, $\color{red}{read-4}$, $\color{red}{write-1^{''}}$, $\color{red}{read-2}^{''}$, $\color{red}{write-1^{'''}}$) are performed every $k$ iterations, where $k$ is the number of delay steps.


\section{Evaluation}
\label{sec:eval_set}
In this section, we first provide warm-up sensitivity analysis of SSD-SGD and present its convergence accuracy compared to SSGD under different delay steps, then evaluate the effectiveness of GLU in maintaining convergence accuracy. Finally, we demonstrate the performance improvement under different configurations and models.

\subsection{Methodology}
\noindent{\textbf{Testbed Setup:}}
Our testbed has 4 physical machines, each with 40 CPU cores, 256GB memory, 4 Tesla V100 GPUs without NVLinks, interconnected with InfiniBand (56Gbps) network. 

\noindent{\textbf{Benchmarks and Datasets:}}
We choose ResNet-20, ResNet-50~\cite{he2016deep}, VGG-11~\cite{simonyan2014very} and AlexNet~\cite{krizhevsky2012imagenet} as our benchmark models,
and CIFAR10~\cite{CIFAR-10} and ImageNet ILSVRC2012~\cite{deng2009imagenet} as the datasets in our experiments. 

\noindent{\textbf{Baselines:}}
We use the vanilla ML frameworks under SSGD as baselines. We also present the linear scalability, which is used as an optimal case in many works~\cite{p3, sergeev2018horovod, zhang2017poseidon}. It is calculated by the training speed on 1 machine (with a vanilla ML framework) multiplied by the number of machines. We use training speed (images/sec) as the performance metric. All the reported speed numbers are averaged over 4 training epochs and test accuracy mentioned in this paper means TOP-1 accuracy.

\noindent{\textbf{Hyper-parameters Setting:}}
GLU algorithm is applied in local update and we need to set parameters $loc\_lr$, $\alpha$ and $\beta$ according to Equation~\ref{eqn:glu-update}. After performing grid search for the hyper-parameters and our 4-machines cluster obtains the best test performances by choosing $\alpha = 2.0$, $\beta = 0.5$ and $loc\_lr$ is 4 times the value of global $lr$ in the parameter servers. For instance, if the global $lr$ in the parameter servers is 0.4, then $loc\_lr$ is set to 1.6. The \textit{WP} stage is 0 by default.

\subsection{Warm-up Sensitivity Analysis}
\label{sec:warm-up sensitivity}
\begin{figure}[!t]
	\centering
	\begin{minipage}{0.46\linewidth}
		\centerline{\includegraphics[width=2.4 in, height= 1.6 in]{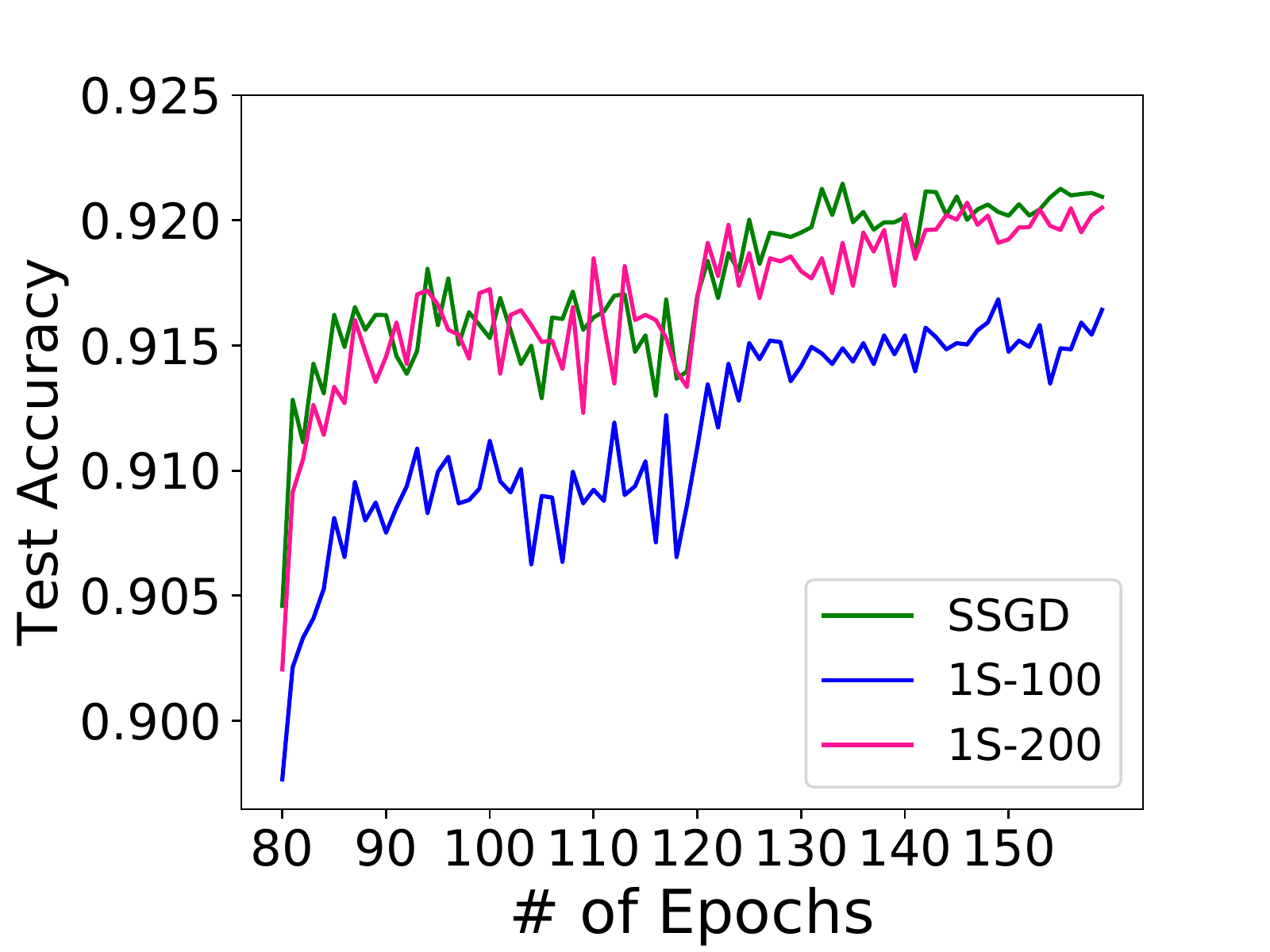}}
	\end{minipage}
	\begin{minipage}{0.46\linewidth}
		\centerline{\includegraphics[width=2.4 in, height= 1.6 in]{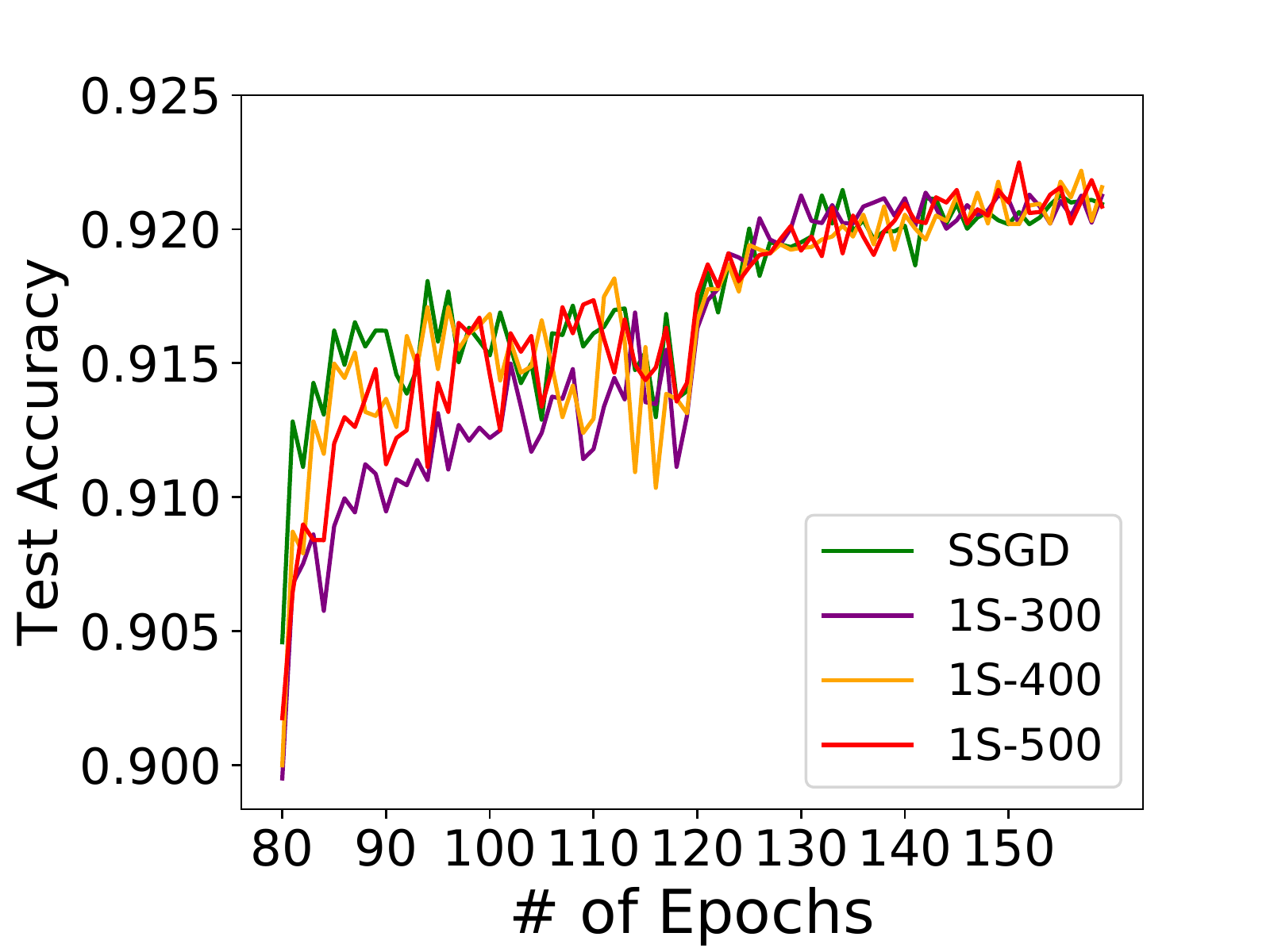}}
	\end{minipage}
	\caption{Test accuracy of ResNet-20 (CIFAR-10, with data augmentation) under different number of warm-up iterations. $1S-100$ means warm up for 100 iterations under the 1 step delay mechanism.}
	\label{fig:warm-up}
\end{figure}

According to discussion in Section~\ref{sec:glu-algorithm}, it is necessary to introduce warm-up stage before the application of steps delay mechanism. 
Fig.~\ref{fig:warm-up} illustrates the test accuracy under different number of warm up iterations, from which we have the following observations: (1) When the warm-up stage is 100 iterations, the test accuracy is obviously lower than that of SSGD. (2) When the warm-up stage is 200 iterations, test accuracy is slightly lower than that of SSGD while much better than $1S-100$. (3) When the warm-up stage is 300 or more iterations, test accuracy of SSD-SGD is comparable to that of SSGD, or even slightly better, and $1S-500$ achieves the best accuracy.
Experimental results proves that warm-up stage is necessary for SSD-SGD, and the period of warm-up stage is set to 500 iterations in all the following experiments.

\begin{table}[b]\scriptsize
	\renewcommand{\arraystretch}{1.3}
	\caption{Test accuracy (\%) of ResNet-20 and ResNet-50.}
	\vspace{-0.5em}
	\label{tab:best-acc}
	\centering
	\begin{tabular}{c|c|c|c|c|c|c}		
		\toprule
		{\bf Delay Steps} &{\bf SSGD} & {\bf 1-Step} & 2-Step & 3-Step & 4-Step & 5-Step\\  
		\hline
		ResNet-20 (32) & 92.126 & 92.204 & {\color{red}{92.228}} & 92.153 & 92.148 & 91.129\\
		\hline
		ResNet-50 (64) & 73.650 & 73.668 & 73.648 & {\color{red}{73.701}} & 73.364 & 73.354 \\
		\hline
		ResNet-50 (32) & 73.792 & 73.865 & 73.810 & 74.001 & {\color{red}{74.004}} & 73.745\\ 
		\bottomrule
	\end{tabular}
\end{table}

\subsection{Convergence Accuracy under SSD-SGD}
In this subsection, we validate the convergence accuracy of SSD-SGD under different delay steps to find the maximum $k$ for various models. GLU algorithm is used for local update and Table~\ref{tab:best-acc} displays the test accuracy.


When training ResNet-20 with CIFAR-10, the batch size is 32 per GPU, the global $lr$ is 0.4 and the $loc\_lr$ is 1.6. Actually, training ResNet-20 with CIFAR-10 dataset is quick and efficient, it is unnecessary for large-scale distributed deployment. We conduct experiments on ResNet-20 to prove that SSD-SGD is applicable to neural networks of low complexity. 
Based on the results of ResNet-20 in Table~\ref{tab:best-acc}, we have the following observations: (1) When the delay steps is less than 5, test accuracy of SSD-SGD is slightly better than that of SSGD. (2) The test accuracy drops significantly (0.997\%) when the delay steps is 5. This results from the model's low complexity, which leads to low sparsity, and models with low sparsity are more sensitive to the number of delay steps. (3) The best test accuracy is obtained when the delay steps is 2.



When training ResNet-50 with ImageNet, the batch size is set to 64 and 32 per GPU separately, the global $lr$ are 0.8 and 0.4 while the $loc\_lr$ are 3.2 and 1.6, the \textit{WP} stage for batch size 64 is 5 epochs. According to the results of ResNet-50 in Table~\ref{tab:best-acc}, we have the following observations: (1) SSD-SGD achieves similar or even  better test accuracy as SSGD under different delay steps. (2) When the delay steps is 5, the test accuracy experiences a slightly drop (0.296\% and 0.047\% separately), which is not as obvious as the ResNet-20 (CIFAR-10) model. This is on account of the higher complexity of ResNet-50, leading to its higher sparsity.
(3) Most of the test accuracy achieved with batch size 32 is slightly higher than that of SSGD, while the result of batch size 64 is the opposite. This result is reasonable as a larger batch size contains more training information, leading to more delayed information under the same delay steps. However, we should notice larger batch size means larger computation overhead, more communication overhead can be overlapped in the training process.
(4) The best test accuracy of ResNet-50 (64) and ResNet-50 (32) is obtained with 3 and 4 delay steps respectively. Combining with the best delay steps of ResNet-20, we come to the conclusion that the best number of delay steps is larger for a model with higher sparsity.

Based on the experimental results above, we have the following conclusion: SSD-SGD is applicable to models of low complexity and high complexity, it can achieve similar or even better test accuracy than SSGD, and the maximum $k$ varies for different models.

\subsection{Effectiveness of GLU algorithm}

\begin{figure}[t]
	\centering
	\begin{minipage}{0.5\linewidth}
		\centerline{\includegraphics[width=2.4 in, height= 1.6 in]{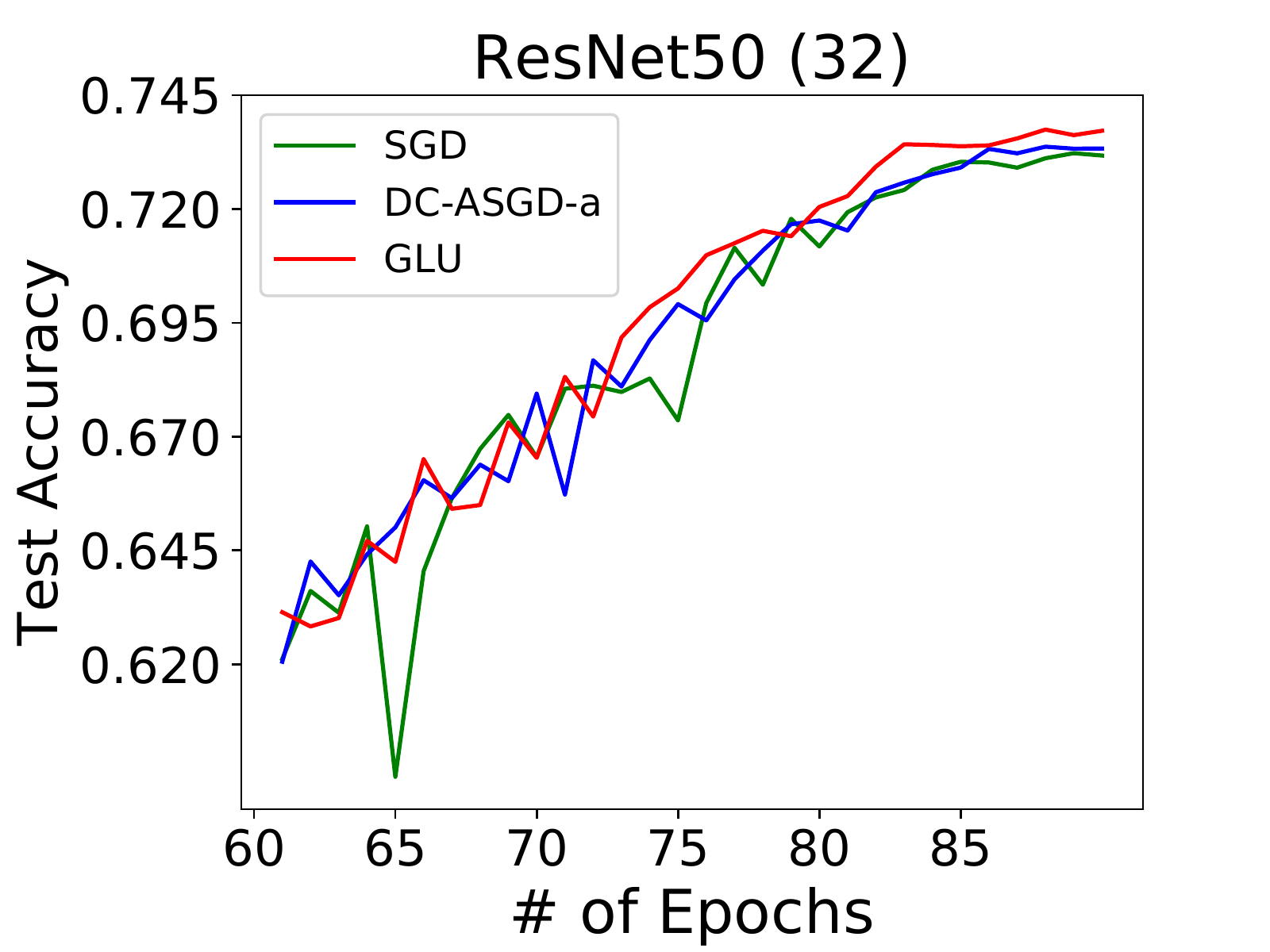}}
	\end{minipage}
	\begin{minipage}{0.4\linewidth}
		\centerline{\includegraphics[width=2.4 in, height= 1.6 in]{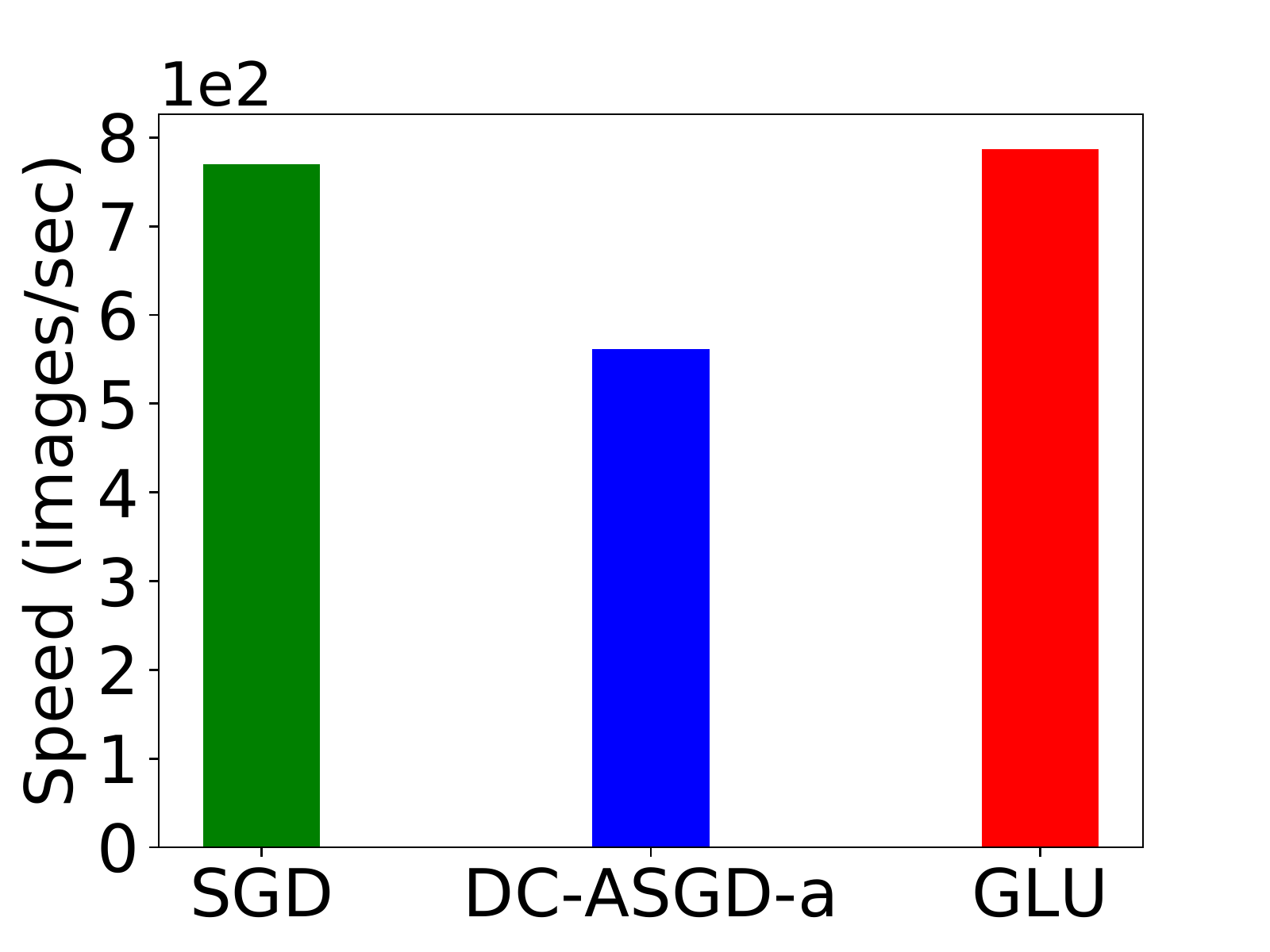}}
	\end{minipage}
	\caption{Test accuracy of ResNet-50 (32) on ImageNet and the corresponding average training speed of one machine under different local update algorithms, the delay steps is 5.}
	\label{fig:glu-prove}
\end{figure}

This subsection is provided to prove the effectiveness of GLU in maintain the convergence accuracy. Fig.~\ref{fig:glu-prove} illustrates the test accuracy and training speed of SSD-SGD under three different local update algorithms, 4 parameter servers and 4 workers are involved in this experiment. The local learning rates for SGD, DC-ASGD-a~\cite{dcasgd-2017} and GLU are 0.1, 0.4, 1.6, respectively, the global learning rates in the servers are 0.4.

According to these two figures, we have the following observations: 
(1) GLU achieves the best test accuracy (73.745\%) and training speed (786.86 images/sec per machine). On the one hand, GLU utilizes global information ($grad_{sync}$) for local update operations. On the other hand, calculation operations introduced by GLU is simple and no operations like matrix multiplication or calculating the square root of matrix are involved. 
(2) Training speed of SGD (769.65 images/sec) is similar to that of GLU while the test accuracy is 0.519\% lower than GLU (73.226\% $vs$ 73.745\%) and 0.143\% lower than DC-ASGD-a (73.226\% $vs$ 73.269\%). This result is reasonable as SGD does not utilize global information, like the $grad_{sync}$ in GLU, for local update operation, making the model converges to a worse accuracy after training the same number of epochs.
(3) DC-ASGD-a obtains worse test accuracy than GLU (73.369\% $vs$ 73.745\%), which can be attributed to two reasons: (a) DC-ASGD-a is proposed to conduct compensation operations in the parameter server while we use it in local workers for local compensations. (b) Configuration of DC-ASGD-a mentioned in~\cite{dcasgd-2017} is for batch size 32 and each GPU is treated as a separate local worker. However, every local worker contains 4 GPUs in our experiments and we linearly increase the learning rate from 0.1 to 0.4, which may not be the optimal configuration. 
(4) DC-ASGD-a has the worst training speed (561.53 images/sec) because of its complex calculation operations. The introduced computation overhead deteriorates the performance improvement of SSD-SGD. This is also the reason why we come up with GLU instead of searching for the optimal configuration for DC-ASGD-a.

\subsection{Speedup under Different Configurations}

\begin{figure}[b]
	\centering
	\begin{minipage}{0.48\linewidth}
		\centerline{\includegraphics[width=2.4 in, height= 1.6 in]{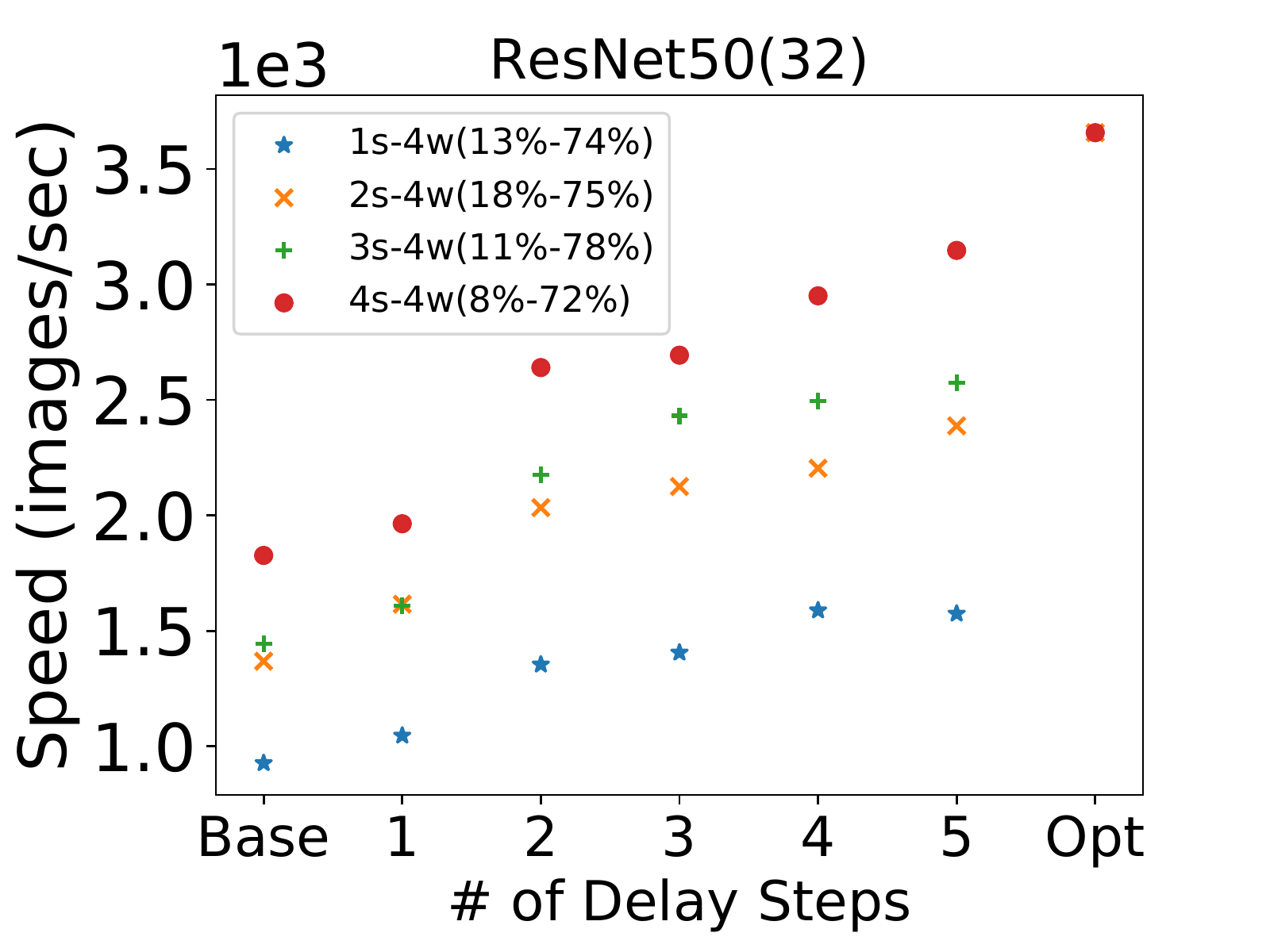}}
	\end{minipage}
	\begin{minipage}{0.48\linewidth}
		\centerline{\includegraphics[width=2.4 in, height= 1.6 in]{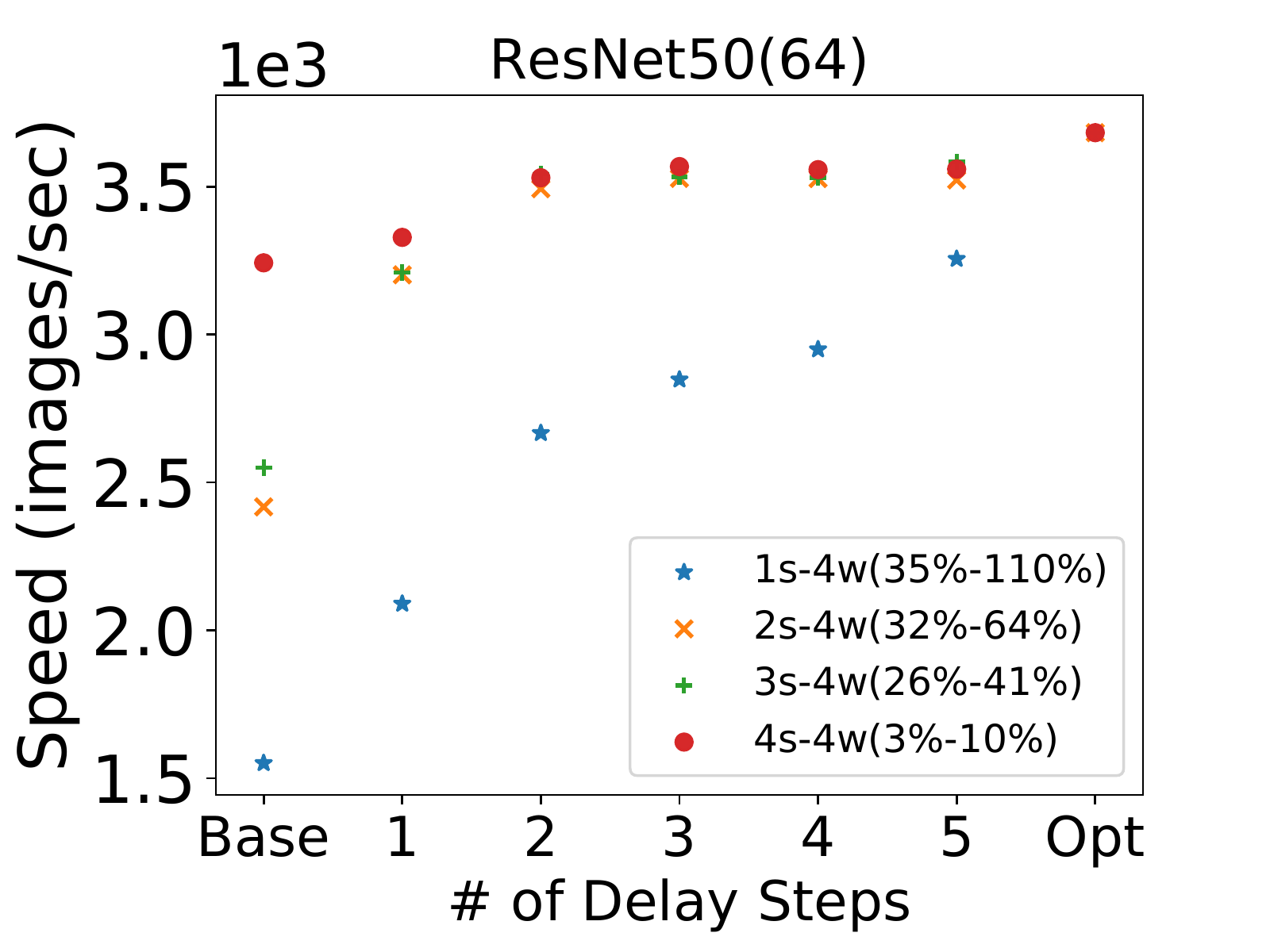}}
	\end{minipage}
	\caption{Training ResNet-50 model with batch size 32 and 64. The numbers in parentheses are SSD-SGD speedup percentages as compared with the baseline. $Base$ means the baseline, $Opt$ means the optimal linear scaling, $4S-4W$ means we set 4 parameter servers and 4 workers.}
	\label{fig:diff-s-w-128-256}
\end{figure}

Fig.~\ref{fig:diff-s-w-128-256} shows the training speeds of baseline (SSGD), SSD-SGD with different delay steps, and linear scaling when training ResNet-50 with the number of parameter servers ranging from 1 to 4, the number of workers is 4.

We adjust the number of servers to adjust the communication overheads in the training process. From Fig.~\ref{fig:diff-s-w-128-256} we have the following observations: (1) SSD-SGD outperforms SSGD by 8\%-78\% across the 5 different delay steps when the batch size is 32, and speedup percentages do not vibrate obviously as the number of servers changes. The bottleneck for further performance improvement lies in the parameter server instead of the communication overhead. The decline in the number of parameter servers requires each server to process more $Push$ and $Pull$ requests per time unit, posing negative impacts on the training speed improvement. 
(2) SSD-SGD outperforms SSGD by 3\%-110\% for batch size 64, and the speedup percentages vibrate greatly as the parameter server number changes. For fixed training epochs, the communication overhead is reduced to half of the original when the batch size is doubled. Therefore, workload of parameter servers is halved by increasing the batch size to 64, which eases the bottleneck for further performance improvement. Speedup percentage for the $1s-4w$ configuration is obviously increased when compared to batch size 32.
(3) It takes 5 delay steps to achieve similar training speed to the optimal case for batch size 32 while only 2 delay steps for batch size 64 under the configuration $4s-4w$, corresponding to 72\% and 10\% speedup respectively. The higher baseline of batch size 64 is attributed to the reduced overhead.
(4) The optimal training speeds of batch size 32 and 64 are similar (3657.22 images/sec $vs$ 3682.12 images/sec). We evaluate the training speed of a single machine (4 GPUs) under different batch sizes, and
the training speed does not go up obviously when the batch size is increased to 32 per GPU.
We attribute this to the limited computing power of GPU, which is close to the peak with batch size 32.

\begin{figure}[t]
	\centering
	\includegraphics[width=2.8in, height= 2in]{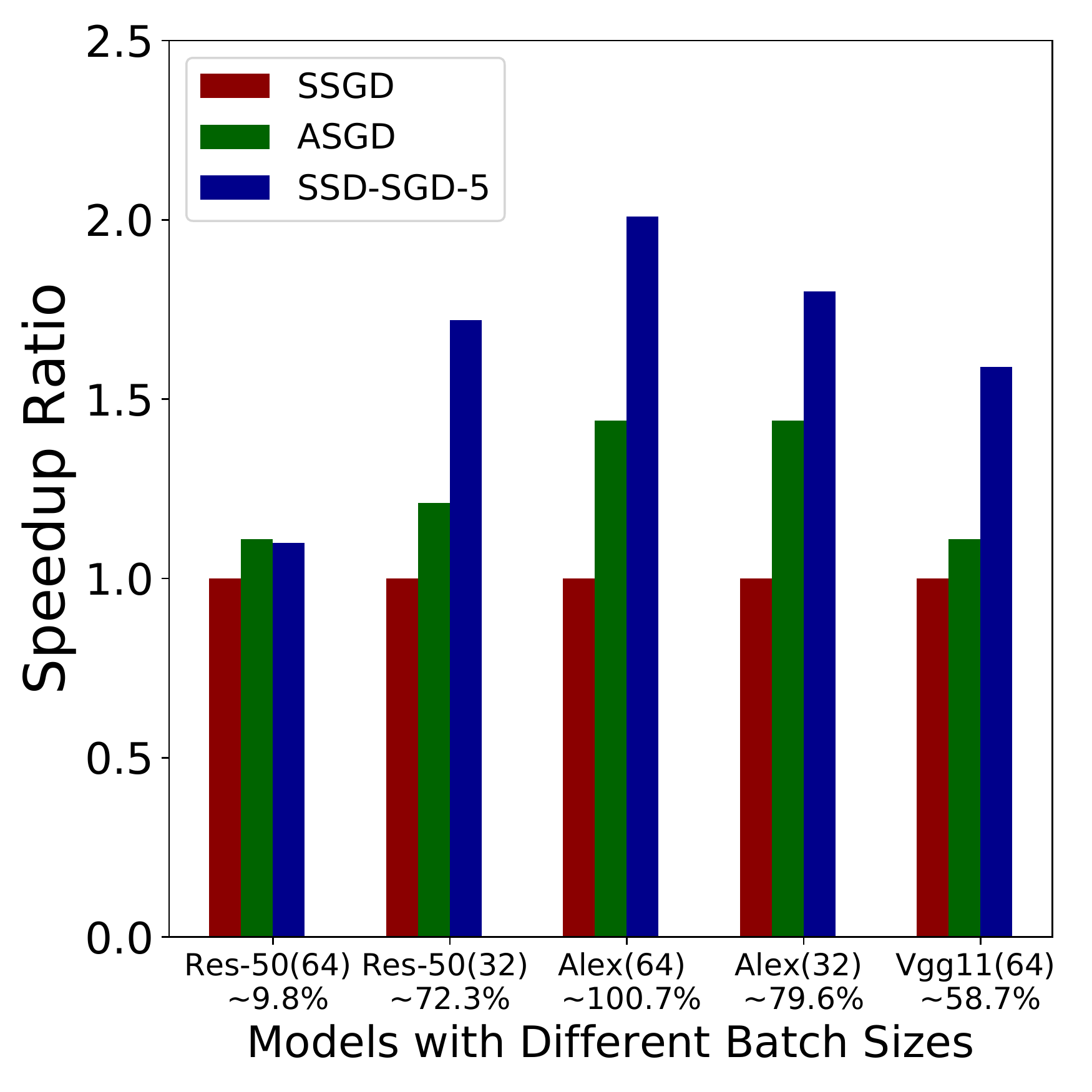}
	\caption{Training ResNet-50, AlexNet and VGG11 models. We normalized the training speed and  numbers under x axis are SSD-SGD-5 speedup percentages as compared with the baseline (SSGD). Res-50(32) means the batch size is 32 per GPU when training ResNet-50.}
	\label{fig:different-models}
\end{figure}

\subsection{Speedup on Different Models}
We also evaluate the speedups of SSD-SGD on communication intensive models like AlexNet and VGG11. Fig.~\ref{fig:different-models} presents the normalized training speed of ResNet-50, AlexNet and VGG11 under SSGD, ASGD and SSD-SGD, and the training speed of SSGD is set to 1. The delay steps of SSD-SGD is 5 (SSD-SGD-5) and 4 parameter servers and 4 workers are involved.

Based on Fig.~\ref{fig:different-models}, we have the following observations:
(1) Training speed of ASGD is slightly faster than SSD-SGD-5 only when training ResNet-50 with batch size 64 per GPU, and two reasons account for its faster speed. On the one hand, ResNet-50 is a computation intensive model and the communication overhead does not significantly slow down the training speed even when training with SSGD. On the other hand, the introduced local update overheads of SSD-SGD-5 imposes a negative influence on the training speed. 
(2) When training ResNet-50 with batch size 32, the communication overhead is doubled for each epoch. The training speed of ASGD deteriorates obviously because of the increased communication overhead, and SSD-SGD-5 obtains 72.3\% speedup.
(3) When training AlexNet with batch size 64, SSD-SGD-5 achieves 100.7\% speedup. As a communication intensive model, the communication overhead of AlexNet makes the bottleneck and training speed can be accelerated greatly by overlapping most of the communication overhead.
(4) SSD-SGD-5 gains 79.6\% speedup when the batch size of AlexNet is reduced to 32. Because the halved computation overhead overlaps only part of the communication overhead under the same delay steps, leading to the lower speedup percentage. However,
ResNet-50 experiences a different trend when the batch size is halved, and this results from its computation intensive feature. 
(5) SSD-SGD-5 achieves 58.7\% speedup on VGG-11 model, which is lower than that on AlexNet. The lower speedup percentage results from the enormous communication overhead of VGG11, whose model size is more than twice that of AlexNet (532.10 MB vs 203.4MB).

This subsection evaluates the effectiveness of SSD-SGD on computation and communication intensive models, and it achieves obvious training speedup on both types of models (9.8\%-100.7\%). Besides, training speed of SSD-SGD-5 is much faster than that of ASGD in most instances.

\section{Related Work}
\label{sec:relate}
Prior studies on communication optimizations can be summarized from the following three terms:

\noindent{\bf Reducing Communication Times:}
Communication times can be reduced with larger batch size, and for fixed epochs, the communication cost will be reduced linearly as the number of communications decreases. Goyal et al.~\cite{goyal2017accurate} firstly increase the batch size to 8K. You et al.~\cite{lars} increase the batch size from 8K to 32K, and Peng Sun et al.~\cite{sun2019optimizing} further increase the batch size to 64K, and so on~\cite{smith2017don, codreanu2017scale, akiba2017extremely, jia2018highly}.

\noindent{\bf Reducing Communication Cost:}
The widely used method for reducing the communication cost for each iteration is cutting down the communication traffic. Gradient compression techniques like Terngrad~\cite{wen2017terngrad}, QSGD ~\cite{alistarh2017qsgd}, GradiVeQ~\cite{yu2018gradiveq} and DGC~\cite{lin2017deep} decrease the communication traffic significantly, while some tricks are needed to get rid of accuracy loss problem. 
BaiDu~\cite{baidu} and IBM~\cite{cho2017powerai} propose efficient all-reduce algorithms for communication, and Horovod~\cite{sergeev2018horovod} introduces the gradient fusion method in all-reduce algorithm to improve the bandwidth utilization. Additionally, communication overhead can be reduced with faster network fabrics~\cite{tpupod, NVIDIA} and efficient network topologies~\cite{bml, dong2020eflops}, too. 


\noindent{\bf Overlap Computation and Communication:}
Apart from the works~\cite{dcasgd-2017, rebutal3-2011distributed, rebutal2-2018stochastic} that use delayed information for training to increase the overlap ratio, a better overlap can also be achieved via communication scheduling techniques.
TicTac~\cite{tictac} and P3~\cite{p3} utilize the prioritization to guarantee near-optimal overlap of communication and computation. 
When compared to TicTac and P3,
ByteScheduler~\cite{bytescheduler} is a more general communication scheduler, which supports TensorFlow, MXNet, PyTorch without modifying their source code. 

\section{Conclusion and Future Work}
\label{sec:conclusion}
SSD-SGD is an algorithm proposed for distributed DNN training acceleration via communication sparsification, which combines the merits of SSGD and ASGD. The core components of SSD-SGD are warm-up stage, steps delay stage and GLU algorithm. We implement SSD-SGD on MXNet framework and evaluate its performance using different models and datasets. Experimental results show that SSD-SGD can obtain similar or even better convergence accuracy than SSGD on models of low and high complexities, and it achieves up to 110\% speedup in training speed. 

As for the future work, we plan to evaluate SSD-SGD on larger computer clusters, where with the increasing number of workers, the delay will become more serious. Furthermore, we will try to improve the GLU algorithm to make better compensation for more serious delay situation.

\bibliographystyle{unsrt}  

\bibliography{reference}
\end{document}